\documentclass[conference]{IEEEtran}
\IEEEoverridecommandlockouts
\usepackage{cite}
\usepackage{amsmath,amssymb,amsfonts}
\usepackage{algorithmic}
\usepackage{graphicx}
\usepackage{textcomp}
\usepackage{xcolor}
\def\BibTeX{{\rm B\kern-.05em{\sc i\kern-.025em b}\kern-.08em
    T\kern-.1667em\lower.7ex\hbox{E}\kern-.125emX}}
    
\usepackage{fancyhdr}
\usepackage{cite}
\usepackage{amsmath,amssymb,amsfonts}
\usepackage{algorithmic}
\usepackage{graphicx}
\usepackage{textcomp}
\usepackage{xcolor}
\usepackage{booktabs}
\usepackage{balance}

\usepackage{soul}
\usepackage{wallpaper}
\usepackage{cleveref}
\usepackage{numprint}
\usepackage{paralist}
\usepackage{xspace}

\usepackage[some]{background}
\usepackage[11pt]{moresize}
\usepackage{t1enc}

\definecolor{orangeN}{rgb}{1,.5,0}
\definecolor{blueN}{rgb}{.2, .59, .88}
\definecolor{purpleN}{rgb}{.294118, 0, .509804}
\definecolor{greenN}{rgb}{.421, .578, .241}
\definecolor{pinkN}{cmyk}{0, 0.7808, 0.4429, 0.1412}
\definecolor{grayN}{gray}{0.6}

\newcommand{\avb}{\textsc{A-VB}~}

\newcommand{\avhigh}{\textsc{A-VB-High}}
\newcommand{\avtwo}{\textsc{A-VB-Two}}
\newcommand{\avcult}{\textsc{A-VB-Culture}}
\newcommand{\avtype}{\textsc{A-VB-Type}}

\newcommand{\humevb}{\textsc{Hume-VB}}

\newcommand{\cmp}{\textsc{ComParE}}
\newcommand{\opensmile}{\textsc{openSMILE}}

\newcommand{\egm}{\textsc{eGeMAPS}}

\newcommand{\eg}{e.\,g., }

\begin{document}

\title{The ACII 2022 Affective Vocal Bursts \\ 
Workshop \& Competition}

\newcommand{\linebreakand}{%
  \end{@IEEEauthorhalign}
  \hfill\mbox{}\par
  \mbox{}\hfill\begin{@IEEEauthorhalign}
}

\author{\IEEEauthorblockN{Alice Baird}
\IEEEauthorblockA{\textit{Hume AI} \\
New York, USA\\
alice@hume.ai}
\and
\IEEEauthorblockN{Panagiotis Tzirakis}
\IEEEauthorblockA{\textit{Hume AI} \\
New York, USA\\
panagiotis@hume.ai}
\vspace{0.4cm}
\IEEEauthorblockN{Anton Batliner}
\IEEEauthorblockA{\textit{University of Augsburg} \\
Augsburg, Germany\\
batliner@uni-a.de}
\and
\IEEEauthorblockN{Jeffrey A. Brooks}
\IEEEauthorblockA{\textit{Hume AI} \\
New York, USA\\
jeff@hume.ai}
\vspace{0.4cm}
\IEEEauthorblockN{Dacher Keltner}
\IEEEauthorblockA{\textit{University of California Berkeley} \\
California, USA\\
keltner@berkeley.edu}
\and 
\IEEEauthorblockN{Chris B. Gregory}
\IEEEauthorblockA{\textit{Hume AI} \\
New York, USA\\
chris@hume.ai} %
\vspace{0.4cm}
\IEEEauthorblockN{Alan Cowen}
\IEEEauthorblockA{\textit{Hume AI} \\
New York, USA\\
alan@hume.ai}
\and
\IEEEauthorblockN{Bj\"{o}rn Schuller}
\IEEEauthorblockA{\textit{Imperial College London} \\
London, UK\\
schuller@ieee.org}

}

\maketitle

\begin{abstract}
The ACII Affective Vocal Bursts Workshop \& Competition is focused on understanding multiple affective dimensions of vocal bursts: laughs, gasps, cries, screams, and many other non-linguistic vocalizations central to the expression of emotion and to human communication more generally. This year's competition comprises four tracks using a large-scale and in-the-wild dataset of 59,299 vocalizations from 1,702 speakers. The first, the \textsc{A-VB-High} task, requires competition participants to perform a multi-label regression on a novel model for emotion, utilizing ten classes of richly annotated emotional expression intensities, including; Awe, Fear, and Surprise. The second, the 
\textsc{A-VB-Two} task, utilizes the more conventional 2-dimensional model for emotion, arousal, and valence. The third, the \textsc{A-VB-Culture} task, requires participants to explore the cultural aspects of the dataset, training native-country dependent models. Finally, for the fourth task, \textsc{A-VB-Type}, participants should recognize the type of vocal burst (e.g., laughter, cry, grunt) as an 8-class classification. This paper describes the four tracks and baseline systems, which use state-of-the-art machine learning methods. The baseline performance for each track is obtained by utilizing an end-to-end deep learning model and is as follows: for \textsc{A-VB-High}, a mean (over the 10-dimensions) Concordance Correlation Coefficient (CCC) of 0.5687 CCC is obtained; for \textsc{A-VB-Two}, a mean (over the 2-dimensions) CCC of 0.5084 is obtained; for \textsc{A-VB-Culture}, a mean CCC from the four cultures of 0.4401 is obtained; and for \textsc{A-VB-Type}, the baseline Unweighted Average Recall (UAR) from the 8-classes is 0.4172 UAR.
\end{abstract}

\section{Introduction}
The \textbf{A}ffective-\textbf{V}ocal \textbf{B}urst (\avb) competition is exploring the expression of affect and emotion in brief nonverbal vocalizations (\eg vocal bursts such as laughs, sighs, and shouts). Within this competition, the organizers provide several emotion modeling strategies and aim to discuss each during the workshop held at the 2022 Affective Computing and Intelligent Interactions (ACII) Conference.   

Thus far, vocal bursts have been largely overlooked in machine learning, affective computing, and emotion science. Given the focus in these fields on facial expressions, the voice has been a relatively understudied medium for communicating emotion. To the extent that the voice has been studied as a modality of emotion expression, it has been chiefly understood from the perspective of speech prosody~\cite{scherer2003vocal}. But another way humans communicate emotion with the voice is with the brief sounds that occur in the absence of speech -- laughs, cries, and shouts (to name a few). Recent studies have discussed the range of emotions conveyed by vocal bursts (known as affect bursts~\cite{scherer1995expression, schroder2003experimental}), with findings demonstrating that brief vocalizations reliably express over ten emotions and that the meanings of vocal bursts are generally preserved across diverse cultures~\cite{cordaro2016voice, cowen2019mapping}.

The field of machine learning has recently seen increased interest in vocal burst modeling, with the Expressive Vocalizations (ExVo) competition at ICML in 2022~\cite{BairdExVo2022} being the first-of-its-kind competition to explore various modeling strategies to understand and generate vocal bursts. More broadly, computational speech-based emotion modeling has become a prevalent area of research since the success of computational paralinguistic methods~\cite{Schuller14-CPE} and general advances in machine and deep learning speech recognition strategies~\cite{amodei2016deep}. Computational modeling of emotion promises to inform a wide range of wellbeing domains, with applications including diagnostic tools for psychiatric illnesses~\cite{mundt2012vocal}, and bio-markers for remote wellness monitoring~\cite{coravos2019developing}.

In the \avb{} competition, 
we extend on our recent works~\cite{BairdExVo2022}, with a more specific focus on comparing and contrasting the various strategies available for modeling emotion in vocal bursts. In particular, the \avb{} competition presents four sub-challenges utilizing a single dataset: \begin{inparaenum}[(1)] \item the high-dimensional emotion task (\avhigh), in which participants must predict a high-dimensional (10 class) emotion space, as a multi-output regression task, \item the two-dimensional emotion task (\avtwo), where the two-dimensional emotion space based on the circumplex model of affect~\cite{russell1980_Circumplex} (arousal and valance) is to be recognized, again as a multi-output regression task, \item the cross-cultural emotion task (\avcult),  where participants will be challenged with predicting the intensity of 10 emotions associated with each vocal burst as a multi-output regression task, using a model or multiple models that generate predictions specific to each of the four cultures provided in the dataset (the U.S., China, Venezuela, or South Africa), and \item the expressive burst-type task (\avtype), in which participants are challenged with classifying the type of expressive vocal burst from 8-classes; \textit{Cry, Gasp, Groan, Grunt, Laugh, Other, Pant, Scream}.  \end{inparaenum}

The dataset used within the \avb{} competition, the Hume Vocal Bursts dataset (\humevb), comprises 59,201 recordings totaling more than 36 hours of audio data from 1,702 speakers. First utilized in the \avb{} competition\cite{BairdExVo2022}, to our knowledge, this dataset remains one of the largest available of vocal bursts. The recordings in \humevb{} are rich and diverse in several ways that present unique opportunities, with the labeling enabling an array of emotion characteristics to be explored from vocal bursts. A single vocal burst can combine classes such as gasps infused with a cry or a scream, ending with a laugh, and offers a vibrant testing bed for emotion understanding and modeling~\cite{cowen2019mapping}. Thus, the \humevb{} dataset enables distinct but complementary strategies: allowing participants to model continuous blends of utterances such as laughs, cries, and gasps as well as the specific meanings of different laughs (amusement, awkwardness, and triumph), cries (distress, horror, and sadness), gasps (awe, excitement, fear, and surprise), and more.

In this paper, we include a description of the \humevb{} dataset in detail (\Cref{sec:data}), provide rules for the four competition tasks (\Cref{sec:tasks}), and present baseline results for each task (\Cref{sec:baselines}). We summarize our results in \Cref{sec:results} and conclude with a discussion of insights from baseline development in \Cref{sec:conc}.

\section{The A-VB Data}
\label{sec:data}

The \avb{} competition relies on the \humevb{} dataset, a large-scale dataset of emotional non-linguistic vocalizations (vocal bursts). This dataset consists of 36\,:47\,:04  (HH\,:MM\,:SS) audio data from 1\,702 speakers aged from 20 to 39 years. The data was gathered in 4 countries with broadly differing cultures: China, South Africa, the U.S., and Venezuela, and individuals are performing emotional mimicry of seed emotion examples. Furthermore,  speakers' are recorded in their homes via their microphones.

Each vocal burst has been labeled in terms of the intensity of 10 different expressed emotions, each on a [1\,:100] scale, and these are averaged over an average of 85.2 raters' responses\begin{inparaenum}[1]\item \textit{Amusement}\item \textit{Awe}\item \textit{Awkwardness}\item \textit{Distress}\item \textit{Excitement}\item \textit{Fear}\item \textit{Horror}\item \textit{Sadness}\item \textit{Surprise}\item and  \textit{Triumph}.\end{inparaenum} 

As well as the distribution of arousal and valance, and cultural-based emotion dimensions of \humevb{}, in \Cref{fig:tsne} (left), a t-SNE representation of emotional expressions based on the human ratings across the training set is visualized. We can see that the expressions vary, with clearly defined regions corresponding to each expressed emotion and continuous gradients between emotions (\eg amusement and excitement). Of note, fewer samples convey \textit{Triumph}, so we expect this class to be more challenging to model.

The intensity ratings for each emotion were scaled to [0\,:1]. For our baseline experiments, the audio files were normalized to -3 decibels and converted to 16\,kHz, 16\,bit, mono (we also provide participants with the raw unprocessed audio, captured at 48\,kHz). The data was subsequently partitioned (see \Cref{tab:splits},) into training, validation, and test splits, considering speaker independence and a balance across classes. 

\begin{figure*}
    \centering
    \includegraphics[width=0.32\linewidth]{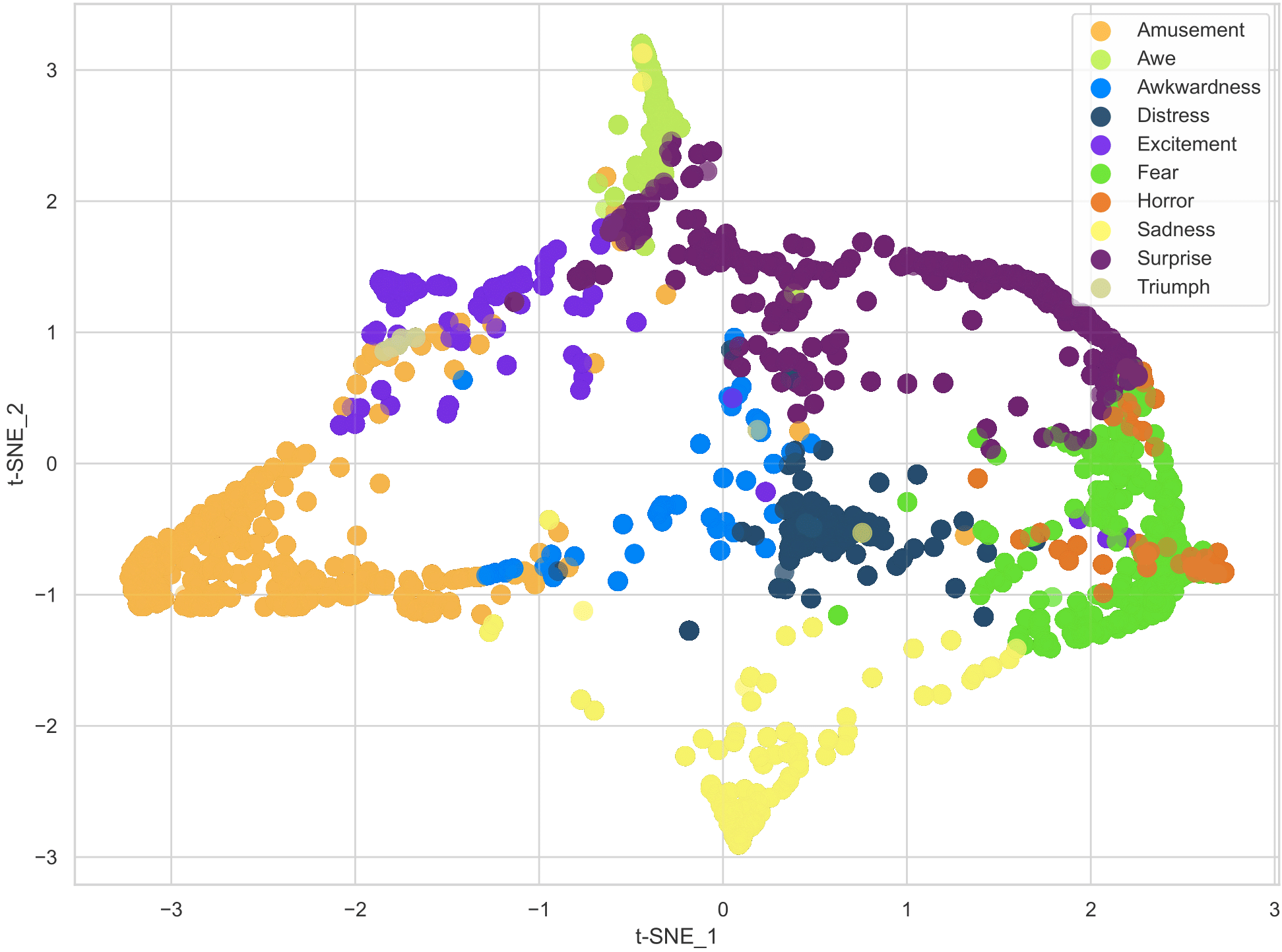}
    \includegraphics[width=0.32\linewidth]{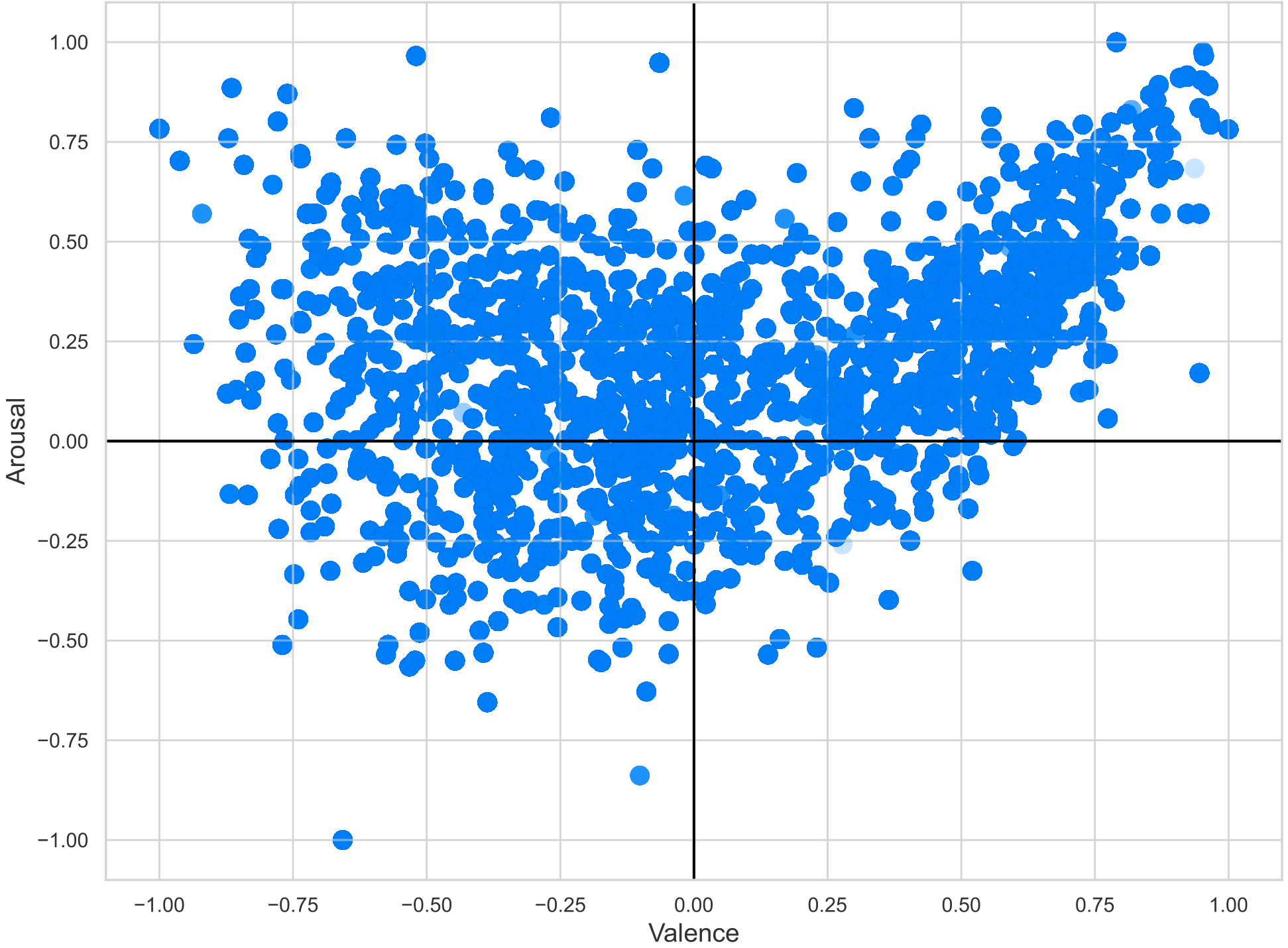}
    \includegraphics[width=0.32\linewidth]{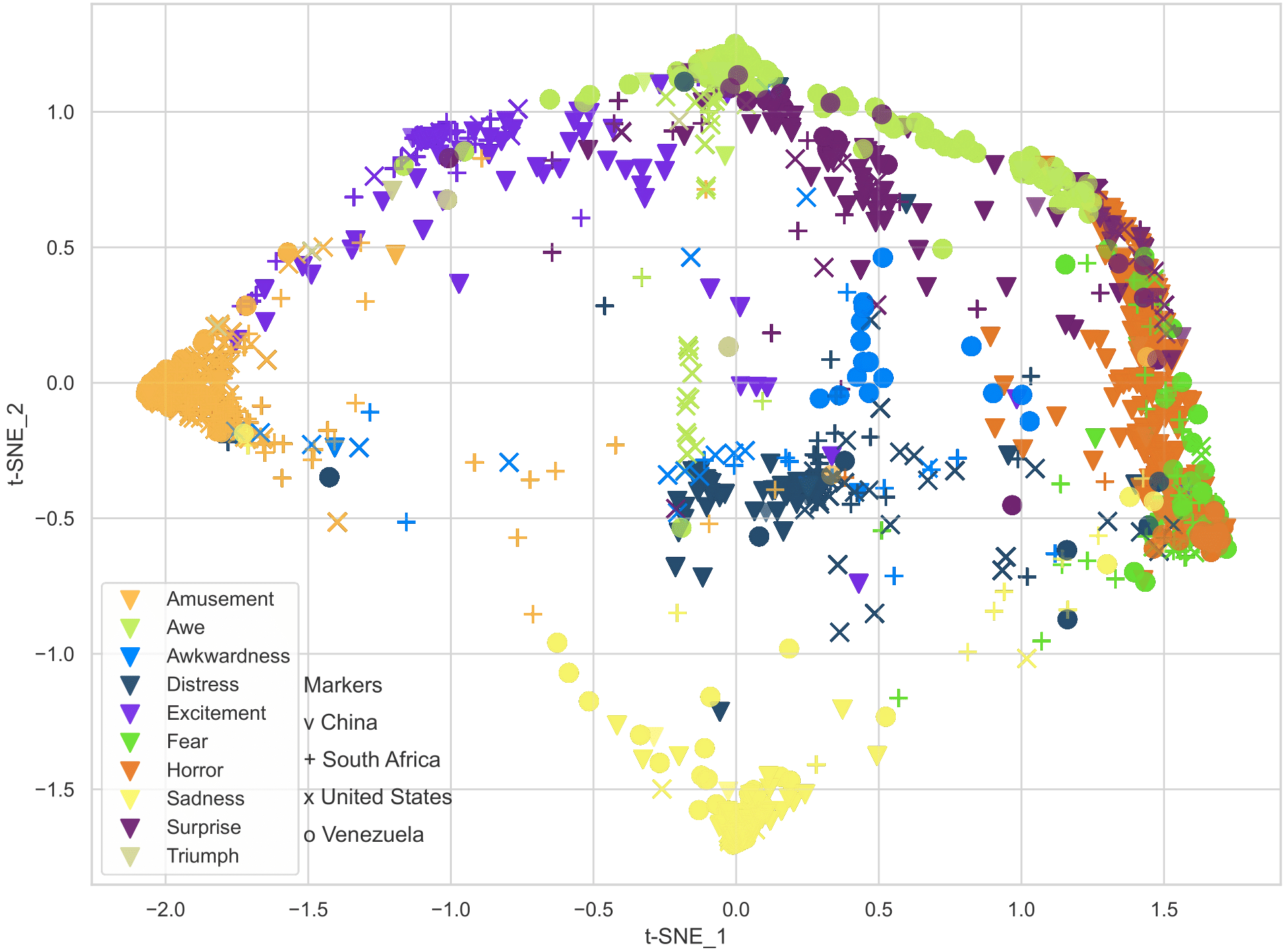}
   
\caption{t-SNE representation of the emotional expression, labeled based on the most dominantly expressed emotion (left), the distribution of arousal and valance (middle), and a t-SNE representation of the culture-based emotion labels (right), from the \humevb{} training set. }
    \label{fig:tsne}
\end{figure*}

\begin{table}
\small
\centering
\caption{An overview of the \humevb{} competition data. Including (No.) samples, duration (HH:\,MM:\,SS), speakers, and country-of-origin. The age range for speakers is 20.5:39.5 years. Due to the competition setting, the test set for this dataset is blinded.}
\resizebox{0.8\columnwidth}{!}{ 

\begin{tabular}{l | r r r r}
\toprule
             & \textbf{Train}    & \textbf{Val.}      & \textbf{Test}     & $\sum$ \\
             \midrule
\textbf{HH:\,MM:\,SS}     & 12\,:19\,:06 & 12\,:05\,:45 & 12\,:22\,:12 & 36\,:47\,:04            \\
\textbf{No.}      & 19\,990   & 19\,396  & 19\,815   & 59\,201              \\
\midrule
\textbf{Speakers}     &   571    &   568    &   563   & 1\,702             \\
\midrule
\textbf{USA}          &   206    &   206    &   ---    & ---           \\
\textbf{China}        &   79     &   76     &   ---    & ---          \\
\textbf{South Africa} &   244    &   244    &   ---    & ---        \\
\textbf{Venezuela}    &   42     &   42     &   ---    & ---            \\
\bottomrule
\end{tabular}
}
\label{tab:splits}
\end{table}

\section{The Competition Tasks}
\label{sec:tasks}

In the \avb competition, we present four tasks 
of varying nature utilizing the \humevb data. Each explores a different aspect of the affective samples, with our aim to understand more deeply the various strategies for modeling emotion in vocalizations -- an ongoing area of research for machine learning. 

\subsection{A-VB High}

In the High-Dimensional Emotion Sub-Challenge (\avhigh), participants are challenged with predicting the intensity of 10 emotions (Awe, Awkwardness, Amusement, Distress, Excitement, Fear, Horror, Sadness, Surprise, and Triumph) associated with each vocal burst as a multi-output regression task. Participants will report the mean Concordance Correlation Coefficient (CCC) across all ten emotions.

\subsection{A-VB Two}

In the Two-Dimensional Sub-Challenge (\avtwo), participants predict values of arousal and valence (based on 1=unpleasant/subdued, 5=neutral, 9=pleasant/stimulated), derived from the circumplex model for affect~\cite{russell1980circumplex} as a regression task. Participants will report the mean CCC across the two dimensions. In \Cref{fig:tsne} (middle) we see the distribution of the valence and arousal ratings a t-SNE representation, showing a broad distribution for valence. 

\subsection{A-VB Culture}

The Cross-Cultural High-Dimensional Emotion Sub-Challenge (\avcult) is a 10-dimensional, 4-country culture-specific emotion intensity regression task. In \avcult{}, participants are challenged with predicting the intensity of 40 emotions (10 from each culture) as a multi-output regression task. The label for each vocal burst consists of a culture-specific gold standard created from the average of annotations from the sample's culture. Participants will report the mean CCC across all 40 emotions. In \Cref{fig:tsne} (right) a t-SNE representation of the per-culture emotions for each country is plotted, showing clusters of each emotion similar to the \avhigh{} task. 

\subsection{A-VB Type}

In the Expressive Burst-Type Sub-Challenge (\avtype), participants are challenged with classifying the type of expressive vocal burst from 8 classes (Gasp, Laugh, Cry, Scream, Grunt, Groan, Pant, Other). Participants will report the Unweighted Average Recall (UAR) as a measure of accuracy.

\begin{table*}[]
\centering
\caption{Baseline scores for \avb 2022. Reporting the mean Concordance Correlation Coefficient (CCC) for the three regression tasks and the Unweighted Average Recall (UAR) across the 8-classes (chance level .125) for \avtype. For each task, the best score on the test set is emphasized as the official baseline. We report the best scores from 5 seeds.}
\resizebox{0.5\linewidth}{!}{
\begin{tabular}{l | r r  r r  r r | r r}
\toprule
Approach & \multicolumn{6}{c|}{CCC} & \multicolumn{2}{c}{UAR}\\
  & \multicolumn{2}{c}{\avhigh} & \multicolumn{2}{c}{\avtwo} & \multicolumn{2}{c|}{\avcult} & \multicolumn{2}{c}{\avtype} \\
        & Val.   & Test      & Val.       & Test      & Val.   & Test      & Val.       & Test      \\
\midrule
\cmp    &  .5154        &  .5214                 &         .4942    &  .4986              &    .3867        &  .3887                   &    .3913    &   .3839      \\
\egm    & .4484    & .4496            &  .4114     &   .4143               &  .3229   &     .3214            &  .3608       &   .3546   \\
\midrule
\textsc{End2You}     & .5638    & \textbf{.5686 }  & .4988    &\textbf{.5084}    & .4359       &  \textbf{.4401 }     &    .4166     &  \textbf{.4172}    \\
\bottomrule
\end{tabular}
\label{tab:results}
}
\end{table*}

\subsection{General Guidelines}
To participate in the \avb{} 2022 competition, all participants are asked to provide a completed copy of the \humevb{} End-User License Agreement (EULA) (more details can be found on the competition homepage\footnote{http://competitions.hume.ai/avb2022}). Participants should submit a paper that meets the official ACII guidelines, describing their methods. (The \avb{} workshop also accepts contributions on related topics.) To obtain test scores, participants should submit their test set predictions to the competition organizers (each team can do this up to 5 times). Participants are free to compete in any or all tasks. 

\section{Baseline Experiments}
\label{sec:baselines}

For each sub-challenge of the \avb competition, we provide a baseline system utilizing well-established methods known in audio-based emotion recognition modeling~\cite{Schuller16-TI2,eyben2015geneva,tzirakis2018end2you}. We provide reproducible code supporting each baseline system on GitHub\footnote{http://github.com/HumeAI/competitions/tree/main/A-VB2022}.

\subsection{Feature-based Approach}

We extract two sets of features, each having precedence in related tasks~\cite{Schuller13-TI2,Schmitt16-ATB,stappen2020muse1}. One feature vector is extracted per sample for each feature set. Using the \opensmile~toolkit~\cite{eyben2010opensmile}, we extracted the 6,373-dimensional \cmp~set and the 88-dimensional \egm~ set. The 2016 COMputational PARalinguistics ChallengE (\cmp)~\cite{schuller2016INTERSPEECH} set contains 6,373 static features computed based on functionals from low-level descriptors (LLDs)~\cite{Eyben13-RDI,Schuller13-TI2}. The extended Geneva Minimalistic Acoustic Parameter Set (\egm)~\cite{eyben2015geneva}, is smaller in size (88-dimensions), and designed for affective-based computational paralinguistic tasks.


We apply a standard neural network (NN) for these experiments. The NN consists of three fully-connected layers, with layer normalization between each and a leaky rectified linear unit (Leaky ReLU) as the activation function. For the regression experiments, sigmoid is applied to the output layer. The loss for each task 
is varied, with multi-label emotion experiments utilizing a combined Mean Square Error (MSE) loss and the classification tasks applying cross-entropy loss, including softmax on the output layer. From several experiments for each task, a global learning rate ($lr$) and batch size ($bs$) is chosen of $lr=10^{-3}$ and $bs=8$. We also apply early stopping (patience of 5, maximum of 25 epochs) to avoid the effects of overfitting the model.

\subsection{End-to-End Approach}

For our end-to-end baseline, we use the multimodal profiling toolkit \textsc{End2You}~\cite{tzirakis2018end2you}. The baseline model comprises a convolutional neural network (CNN) that extracts features from each audio frame and a recurrent neural network (RNN) that extracts temporal features. We use the Emo-18 (CNN) network architecture~\cite{tzirakis2018end}, which consists of three cascade blocks of 1-D CNN layers, a Leaky ReLU activation function ($\alpha=0.1$), and max-pooling operations. Both convolution and pooling operations are performed in the time domain, using the raw waveform as input. Before the final emotion prediction, we exploit temporal patterns in the signals using a 2-layer Long-Short Term Memory (LSTM) network.

The input audio frame passed to the CNN is $0.1$\,sec long, corresponding to a $1\,600$ dimensional vector, corresponding to the audio sampling rate of 16\,kHz. Audio signals with a length not divisible by the input length are padded with zeros.

Our model is trained with the Adam optimization algorithm~\cite{adam}, a batch size of $8$, and an initial learning rate of $10^{-4}$. The network weights have been initialized with Kaiming uniform~\cite{he2015delving} initialization, and the biases are initially set to zero. The LSTM network comprises 256 hidden units and is trained with a gradient norm clipping of $5.0$. Finally, we use the MSE loss function and the CCC evaluation metric for the regression tasks. For the classification task, we use the cross-entropy loss with UAR as the evaluation metric.

\begin{figure}
    \centering
        \includegraphics[width=0.46\columnwidth]{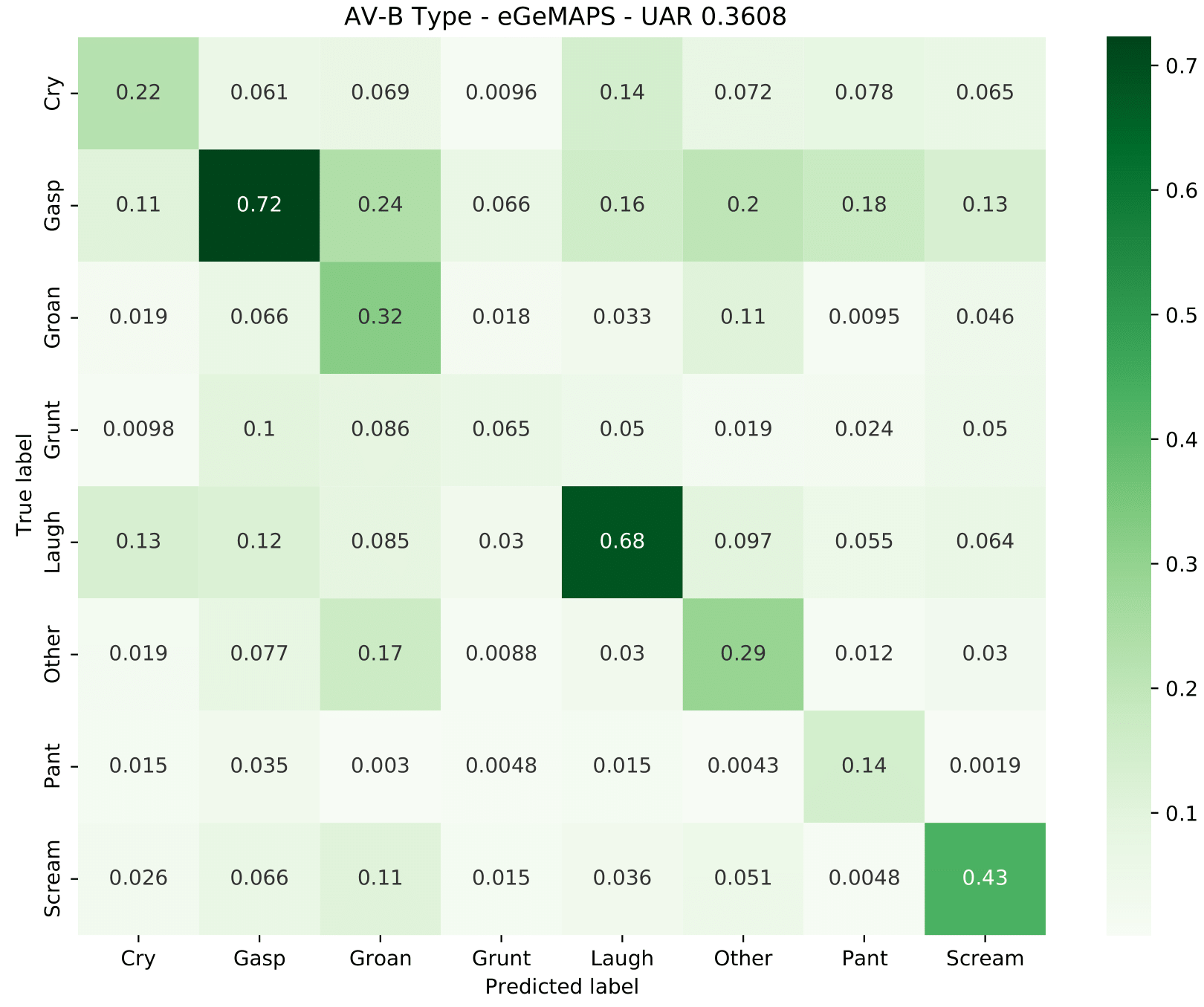}
        \includegraphics[width=0.46\columnwidth]{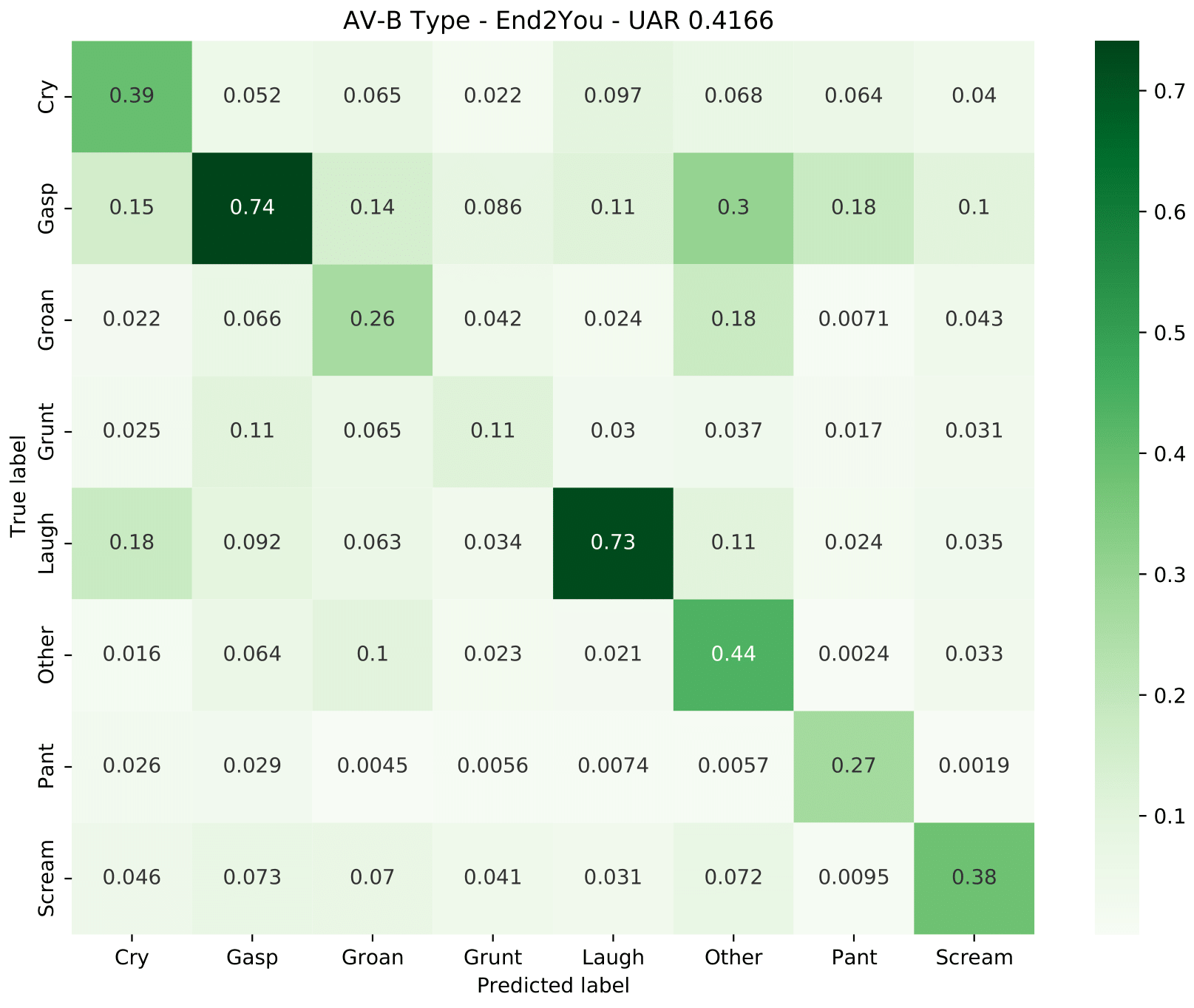}
    \caption{Normalized confusion matrix for validation results of \avtype, with \egm~(left) and  \textsc{End2You} (right) approaches.}
    \label{fig:cm}
\end{figure}

\section{Discussion of Competition Baselines}
\label{sec:results}

In \Cref{tab:results}, we provide the baseline results for each of the four sub-challenges of the \avb competition. In all cases, the baseline score is set by the end-to-end approach \textsc{End2You}, with feature-based strategies falling short in all cases. 

For the \avhigh{} task, a baseline on the test set of 0.5687 CCC is obtained utilizing the end-to-end, \textsc{End2You} method. Of interest here, we see the \cmp{} features closely following much better than \egm. This suggests that the prosodic- and spectral-based features included with the \cmp{} set may benefit this task. On the other hand, the limited samples available may also restrict the potential performance possible from the \textsc{End2You} method. 

We see similar results for \avtwo, with a baseline on the test set of 0.5084 CCC obtained for the mean across the two classes, arousal, and valance. Of interest, we find that the score for valance is higher than for arousal, 0.5701 and 0.4468 CCC, respectively. Typically, arousal would be easier than valance to model from speech~\cite{schuller2020INTERSPEECH}. However,  arousal tends to correlate highly with traits including speech-rate~\cite{hecker1968manifestations}, and volume~\cite{hendrick1970effects}. With this in mind, this data is non-language based, and we consider that arousal may be more of a challenge in this context, as these samples are mainly single bursts, and volume may be less impacting on the perception of arousal given varied recording environments.

As with \avhigh{} and \avtwo{}, 
the baseline is set by the \textsc{End2You} approach for \avcult, with a CCC of 0.4401 CCC on the test set. Given the multi-cultural nature of this task, the overall CCC is lower than the others, as some cultures are more difficult to model. This deficiency is shown for Venezuela (a mean of 0.3888 CCC) and China (a mean of 0.3870 CCC), possibly due to the smaller sample size and the cultural difference in these samples.

For the \avtype{} task, we explore classification for the first time with this data, classifying 8-classes of vocalization types. Once again, the \textsc{End2You} approach is set as the baseline (0.4172 UAR on the test set), with a similar margin to the hand-crafted feature-based methods. In \Cref{fig:cm}, we can see the confusion matrix for the test results of the baseline system and the \egm{} approach. 
The most commonly confused class appears to be `Gasp' in both cases, possibly caused by the class imbalance, given that the `Gasp' class is the most dominant (7,104 samples vs.\ 4,940 for `Laugh', on the training set). Furthermore, we see that the hand-crafted features perform better for some classes, mainly `Screaming' in the case of \egm; this may indicate that the speech-based features are valuable for this task, supported by their strong performance across tasks. 

\section{Concluding Remarks}
\label{sec:conc}

This contribution introduced the guidelines and baseline scores for the first ACII Affective Vocal Bursts (\avb) competition. The competition focuses on strategies for computationally modeling emotion in vocal bursts and utilizes a large-scale dataset, the \humevb{} corpus. In this year's \avb, four tasks are presented: (1) \avhigh{}, a multi-label regression task utilizing 10 dimensions of emotion, we report a baseline score of \textbf{0.5686 CCC for \avhigh}; (2) \avtwo{}, modeling two-dimensions of emotion (arousal and valance), we report a baseline score of \textbf{CCC of 0.5084 for \avtwo{}}; (3) \avcult{} in which participants should model 40 emotional dimensions, 10 for each culture, we report a baseline score of \textbf{0.4401 CCC for \avcult}; and (4) \avtype, an 8-class classification of vocalization type, we report a baselines score of \textbf{0.4172 UAR for \avtype}. Several aspects can be explored by participants of the \avb{} competition to improve on the provided baselines. Namely, exploring the advantages of jointly learning from the various labeling provided and knowledge-based approaches which harness the diversity present in the \humevb{} dataset.

\balance
\bibliographystyle{IEEEtran}
\bibliography{main}

\end{document}